\documentclass[apjl]{emulateapj}
\usepackage{apjfonts}

\shorttitle{HST/NICMOS detection of HR~8799~b in 1998}
\shortauthors{Lafreni\`{e}re et al.}

\newcommand{\msun}{\ensuremath{M_{\odot}}}
\newcommand{\lsun}{\ensuremath{L_{\odot}}}
\newcommand{\mjup}{\ensuremath{M_{\rm Jup}}}

\begin{document}

\title{HST/NICMOS detection of HR~8799~\lowercase{b} in 1998}

\author{David Lafreni\`ere\altaffilmark{1}, Christian Marois\altaffilmark{2}, Ren\'e Doyon\altaffilmark{3}, Travis Barman\altaffilmark{4}}
\altaffiltext{1}{Department of Astronomy and Astrophysics, University of Toronto, 50 St. George Street, Toronto, ON, M5S 3H4, Canada}
\altaffiltext{2}{National Research Council Canada, Herzberg Institute of Astrophysics, 5071 West Saanich Road, Victoria, BC, V9E 2E7, Canada}
\altaffiltext{3}{D\'epartement de physique and Observatoire du Mont M\'egantic, Universit\'e de Montr\'eal, C.P. 6128 Succ. Centre-Ville, Montr\'eal, QC, H3C 3J7, Canada}
\altaffiltext{4}{Lowell Observatory, 1400 West Mars Hill Road, Flagstaff, AZ 86011, USA}

\email{lafreniere@astro.utoronto.ca}

\begin{abstract}
Three planets have been directly imaged around the young star HR~8799. The planets are 5--13~\mjup\ and orbit the star at projected separations of 24--68~AU. While the initial detection occurred in 2007, two of the planets were recovered in a re-analysis of data obtained in 2004. Here we present a detection of the furthest planet of that system, HR~8799~b, in archival HST/NICMOS data from 1998. The detection was made using the locally-optimized combination of images algorithm to construct, from a large set of HST/NICMOS images of different stars taken from the archive, an optimized reference point-spread function image used to subtract the light of the primary star from the images of HR~8799. This new approach improves the sensitivity to planets at small separations by a factor of $\sim$10 compared to traditional roll deconvolution. The new detection provides an astrometry point 10 years before the most recent observations, and is consistent with a Keplerian circular orbit with $a\sim$70~AU and low orbital inclination. The new photometry point, in the F160W filter, is in good agreement with an atmosphere model with intermediate clouds and vertical stratification, and thus suggests the presence of significant water absorption in the planet's atmosphere. The success of the new approach used here highlights a path for the search and characterization of exoplanets with future space telescopes, such as the James Webb Space Telescope or a Terrestrial Planet Finder.
\end{abstract} 
\keywords{planetary systems --- techniques: image processing}

\section{Introduction}

After more than a decade marked by the great success of the radial velocity and transit planet detection techniques, the direct imaging technique has finally made its grand entry on the scene in late 2008 through a series of exciting exoplanets discoveries \citep{marois08,kalas08,lafreniere08,lagrange08}, including the spectacular discovery of the multiple-planet system HR~8799. This system showcases three planets of mass 5--13~\mjup\ orbiting at projected separations of $\sim$24, 38 and 68~AU. It is located 39.4~pc away from the Sun and appears to be seen nearly face-on. The planets all orbit the star in the same direction and likely in the same plane, consistent with formation within a circumstellar disk. The host is a 30--160~Myr-old star of spectral type A5, mass $\sim$1.5~\msun\ and luminosity $\sim$4.9~\lsun; it is also classified as both $\gamma$~Dor and $\lambda$~Boo types. A large infrared excess is detected at $\sim$100~$\mu$m, suggesting that the planetary system is surrounded by a large dust disk. The reader is referred to \citet{marois08} and references therein for further details on this system.

As the first multiple-planet system directly imaged, HR~8799 offers many new possibilities to advance our understanding of planets. First, multi-wavelength photometry and spectroscopy of these three planets, which are almost certainly of the same age and metallicity, will be of great value for the validation and calibration of both evolutionary and atmospheric models of giant planets. Also, with good measurements of the orbits of the planets, it could be possible to independently constrain their masses through dynamical studies, as already attempted by \citet{fabrycky08}. 

Besides its high scientific interest, the discovery of the HR~8799 system represents a great technical achievement given the brightness ratios and angular separations between the planets and the star. The main obstacle to direct imaging of exoplanets, for both space- and ground-based telescopes, is the scattering of stellar light by irregularities on optical surfaces which creates a halo of bright quasi-static speckles around the stellar core, masking the underlying fainter planets. The planets around HR~8799 were initially discovered using adaptive optics and the angular differential imaging (ADI) technique \citep{marois06}, which uses the natural field-of-view rotation of an alt-az telescope to discriminate planets from quasi-static speckles. Effectively, this techniques allows to construct a high-fidelity image of the stellar point-spread function (PSF) that does not contain the signal of any eventual planets; this reference PSF image is subtracted from the target image to remove the light from the star while preserving that of any eventual planet. This technique, coupled with the locally-optimized combination of images (LOCI) algorithm \citep{lafreniere07a}, has proved to be the most efficient approach to detect planets from the ground \citep[see e.g.][]{lafreniere07b}.

Following the initial detection of HR~8799~b and c in October 2007, a careful re-analysis of adaptive optics observations of HR~8799 obtained in July 2004, using improved PSF subtraction algorithms, allowed to recover both planets, thus providing a baseline of four years relative to the most recent observations of September 2008. This four-year baseline firmly establishes common proper-motion of the two outermost planets and clearly reveals their orbital motion. However, this baseline is still quite small compared to their orbital periods ($P>200$~yr), and astrometric measurements over a longer baseline would be highly valuable in better constraining their orbits, and consequently the star and planets masses. This will require several years of observations, but could be achieved more rapidly should a similar re-analysis be applied successfully to even earlier data.

HR~8799 was observed with HST/NICMOS in 1998 as part of a direct imaging survey for massive planets around young nearby stars (program 7226, PI Eric Becklin); the main results of this survey were published in \citet{lowrance05}. Although these observations have, in principle, sufficient angular resolution and flux sensitivity to see the planets of the HR~8799 system, the scattered light from the bright primary prevented their actual detection. This scattered light can be partly removed by subtracting images obtained at two different spacecraft roll angles, the so-called roll-deconvolution technique \citep{schneider03}, but this is insufficient to reveal the planets because of important PSF evolution between the two roll angles. As part of the same HST program, many other stars were observed using the same instrumental configuration as for HR~8799, and as a result several images display PSFs very similar to that of HR~8799. As suggested in \citet{lafreniere07a}, such observations provide a very interesting basis for constructing an optimized reference PSF image using the LOCI algorithm. The LOCI algorithm, detailed in \citet{lafreniere07a}, determines the coefficients needed to linearly combine several reference images into an optimized reference PSF image whose subtraction from the target image will minimize the residual noise. An important and powerful feature of this algorithm is its flexibility to optimize the PSF subtraction locally over several sub-sections of the image. Applied to the HST data set mentioned above, the LOCI algorithm could potentially reduce the scattered light of HR~8799 sufficiently to reveal the planets. With this in mind we have re-analyzed these data and here we report the successful detection of the furthest of the three planets, HR~8799~b, thus extending astrometric measurements of this planet to ten years. Incidentally, this means that an exoplanet could have been discovered by direct imaging only a few years after the discovery of 51~Peg~b by the radial velocity technique.
 
\section{Data set and analysis}\label{sect:analysis}

All the data used in this study come from HST program 7226, which was a survey for giant planets and brown dwarfs around young nearby stars. The observations used the NICMOS medium resolution camera NIC2, the F160W filter and the 0.63\arcsec\ diameter occulting spot to reduce the diffracted light from the primary star. Typically three images of each target were obtained at each of two roll angles differing by 29.9\degr. A total of 39 targets were observed for this program. The pipeline-reduced images used here were retrieved from the HST archive housed at the CADC.\footnote{\url{http://www2.cadc-ccda.hia-iha.nrc-cnrc.gc.ca/cadc/}}

\begin{figure}
\epsscale{1}
\plotone{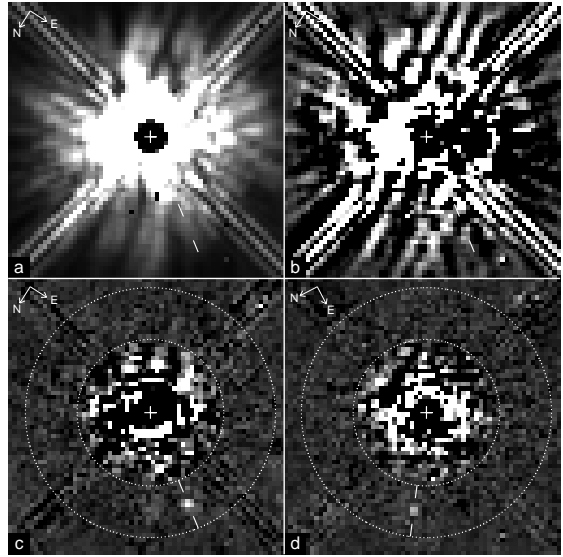}
\caption{\label{fig:im} ({\it a}) Image of HR~8799 at first roll angle. ({\it b}) Image at the first roll angle minus image at the second roll angle. ({\it c} and {\it d}) Residual images, at both roll angles, after subtraction of an optimized reference PSF image obtained using the LOCI algorithm, see text for more details. The display is from 0 to 100 counts s$^{-1}$ for panel {\it a} and from -0.5 to 2.5 counts s$^{-1}$ for panels {\it b--d}. Each panel is 4.8\arcsec\ on a side. The dotted circles in the bottom two panels mark the inner and outer limits of the optimization annulus used for the LOCI algorithm. The two short radial lines enclose the position of HR~8799~b. To produce the images shown in panels {\it c} and {\it d}, a second optimization was performed for the half annulus not containing HR~8799~b, such that the full annulus is representative of the residual noise.}
\end{figure}

As mentioned above, the LOCI algorithm was used with this large set of well-correlated images to construct and subtract an optimized PSF for each of the six images of HR~8799. A priori, the set of reference images used by the LOCI algorithm could include all images of targets that are not HR~8799, but in practice we have omitted some targets from the set because they showed a second point source within a separation of 3\arcsec\ from the star, were significantly fainter than HR~8799, or were not centered behind the coronagraph. The final set of reference images included 203 images from 23 unique sources. When analyzing a given image of HR~8799, we have added to this set the three images of HR~8799 obtained at the other roll angle. Before execution of the LOCI algorithm, all reference images were registered to the target image based on a cross-correlation of the secondary mirror support diffraction spikes. The algorithm was first applied with the specific goal of detecting HR~8799~b, and thus a single optimization region was used. This optimization region was defined as a half annulus extending radially from 1.275\arcsec\ to 2.175\arcsec\ and centered on, but excluding pixels within a radius of 5 pixels from the calculated position of HR~8799~b. The exclusion was used to prevent the algorithm from trying to subtract the companion itself, thus biasing the determination of the coefficients of the linear combination and affecting the residual flux of the companion. The area of the optimization region, $\sim$900 pixels, is equivalent to 225 PSF cores (i.e. the characteristic size of a speckle). This number of ``degrees of freedom'' in the optimization region may be compared with the number of different variables, 46 different observations (23 sources each at two roll angles), in the linear combination of the optimum reference image. The different observations do not represent free parameters of the model PSF, however, as they are highly correlated with each other.

The companion HR~8799~b was detected in all six residual images and its position between the two rolls angles is precisely equal to the spacecraft rotation angle applied. The three residual images at each roll angle were co-added and the results are shown in Fig.~\ref{fig:im}. At the peak pixel of the companion in these co-added images, the detections are at the levels of 10$\sigma$ and 6$\sigma$ for the first and second roll angles, respectively. We have verified that the residual noise in the optimization region closely follows a Gaussian distribution, thus the significance of the detection is high. The improvement in PSF subtraction provided by the LOCI algorithm is obvious from Fig.~\ref{fig:im}, in which the much higher residuals of the roll-subtracted image is apparent. At the separation of HR~8799~b, the residual noise in the LOCI-subtracted image is nine times lower than for the classical roll-deconvolution image (panel {\it b} of Fig.~\ref{fig:im}). It is interesting to note that the LOCI subtraction almost reaches the PSF photon noise. At the separation of HR~8799~b, we estimate that the total residual noise is $\sim$1.8 times more important than the local photon noise.

To confirm that the above detection is not an artifact from the image processing done, we have repeated the analysis 1) without excluding the region containing the companion from the optimization half-annulus, and 2) using only half of the available reference images. The companion was still well detected in both cases, although at lower signal-to-noise ratios (S/Ns) as expected. We have repeated the above analysis with an optimization region designed specifically for HR~8799~c, at $\sim$0.9\arcsec, but the results were inconclusive. The same applies for d, which is at an even smaller separation.

\section{Results}\label{sect:results}

The astrometric and photometric analysis of the companion was done using model PSFs generated with the TinyTim\footnote{\url{http://www.stsci.edu/software/tinytim/tinytim.html}} software (version 7.0)\citep{krist93}. Model PSFs appropriate for the NIC2 pre-cryocooler camera with the F160W filter were generated for each of three approximate positions on the detector: the central star, the companion at the first roll angle, and the companion at the second roll angle. All model PSFs were generated for a width of 10\arcsec\ and a pixel oversampling factor of 9 to minimize interpolation effects when shifting images. Accordingly, during the analysis all shifts were done on the oversampled images, which were then binned $9\times9$ pixels to match the actual NIC2 plate scale. Initially, the model PSFs are normalized to a total flux of one.

For each image, the appropriate companion model PSF was first spatially shifted and intensity scaled over a grid in $dx$, $dy$, and flux. Then each model PSF in this 3D parameter space was subtracted from the image and the residual noise in a $7\times7$-pixel box centered on the companion was computed. The combination of $dx$, $dy$, and flux that yielded the minimum residual noise was found and provided the coordinates and flux of the companion. The uncertainty on the position of the companion, 5~mas, was estimated from the dispersion of the measurements made in all individual images. The center of the star was determined by shifting the appropriate model PSF to maximize the cross-correlation of its diffraction spikes with those of the actual image, as well as by a cross-correlation of the observed PSF diffraction spikes with themselves after a rotation of 180\degr\ about the inferred center; the maximum difference between these two centroid measurements, $\sim$0.1 pixel or 7.5~mas, was taken as an estimate of the centroid accuracy. The pixel coordinates of the star and companion were converted to RA and DEC using the astrometry solution defined in the image FITS headers. A S/N-weighed mean of the relative position of HR~8799~b over the six images was finally calculated; the results are indicated in Table~\ref{tbl:pos}.

\begin{deluxetable}{lcccc}
\tablewidth{230pt}
\tablecolumns{5}
\tablecaption{HR~8799~\rm{b} astrometric measurements \label{tbl:pos}}
\tablehead{
\colhead{Epoch} & \colhead{$\Delta \alpha$ (mas)} & \colhead{$\Delta \delta$ (mas)} & \colhead{Sep (mas)} & \colhead{PA (\degr)}
}
\startdata
1998.8285 & $1411\pm9$ & $986\pm9$ & $1721\pm12$ & $55.1\pm0.4$ \\
2004.5337 & $1471\pm5$ & $884\pm5$ & $1716\pm7$ & $59.00\pm0.23$ \\
2008.6267 & $1527\pm2$ & $800\pm2$ & $1724\pm3$ & $62.35\pm0.09$
\enddata
\tablecomments{Ep. 1998 from this work; others from \citet{marois08}.}
\end{deluxetable}

Since the model PSFs span 10\arcsec, the companion flux found by the above procedure is effectively the same as that in an infinite aperture. The information in the {\em NICMOS Data Handbook v7.0} was then used to convert this flux ($20.5\pm2.5$~counts~s$^{-1}$) to physical units, yielding $(4.2\pm0.5)\times10^{-5}$~Jy, or a magnitude of $18.54\pm0.12$. This magnitude is also indicated in Table~\ref{tbl:phot} along with the measurements reported by \citet{marois08} and the expected magnitudes for two different atmosphere models (see \S\ref{sect:discussion} for more detail on models).

\begin{figure}
\epsscale{1}
\plotone{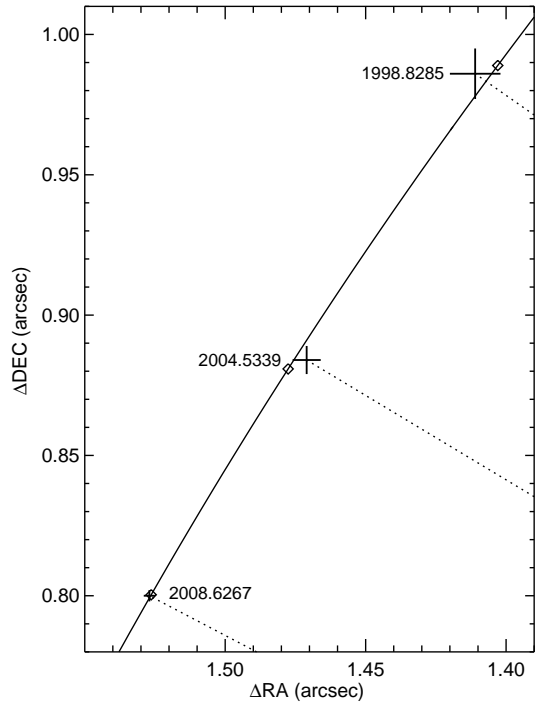}
\caption{\label{fig:pos} Measured relative positions (with associated errors) of HR~8799~b at the three available epochs. The solid curve shows a Keplerian circular orbit with $a=71.4$~AU, $i=19.2\degr$, and position angle of ascending node $\Omega=131.3$\degr; the open diamonds mark the corresponding planet position at the three epochs. The $\chi^2$ for this orbit is 3.19.}
\end{figure}

\begin{deluxetable}{lccc}
\tablewidth{230pt}
\tablecolumns{4}
\tablecaption{HR~8799~\rm{b} photometric measurements \label{tbl:phot}}
\tablehead{
\colhead{Filter} & \multicolumn{3}{c}{Magnitude} \\
\cline{2-4}
& \colhead{Measured} & \colhead{Intermediate clouds} & \colhead{Extreme clouds}
}
\startdata
$J$ & $19.28\pm0.16$ & 19.42 & 19.66  \\
F160W & $18.54\pm0.12$ & 18.58 & 18.31  \\
$H$ & $17.85\pm0.17$ & 18.30 & 18.08 \\
$K_{\rm s}$ & $17.03\pm0.08$ & 17.43 & 16.91  \\
$L^\prime$ & $15.64\pm0.11$ & 15.24 & 15.59
\enddata
\tablecomments{F160W from this work; others from \citet{marois08}.}
\end{deluxetable}

\section{discussion}\label{sect:discussion}

As visible in Fig.~\ref{fig:pos}, the new position is consistent with the previous ones, and still suggests a nearly circular orbit seen close to face-on. We have done very simple orbital fits to verify that the data are consistent with true Keplerian orbits. For simplicity, we have considered only circular orbits and assumed that the stellar mass is precisely equal to 1.5~\msun, and then we explored a range of semimajor axes ($a=60$--100~AU) and inclinations ($i=0$--45\degr). The best fits, with a $\chi^2\sim3.2$, were found for $a\sim68$--74~AU and $i\sim13$--23\degr; an example of such fit is shown in Fig.~\ref{fig:pos}. This range of inclination is in line with what is expected for the equatorial plane of the star based on its measured $v \sin{i}$ (37.5~km~s$^{-1}$, \citealt{gray99}) which, for the range of true rotation velocity of A5 stars (100--300~km~s$^{-1}$, \citealt{royer07}), would yield $i=7\degr$--22\degr. We refrain from carrying out more detailed orbital fits as the limited time baseline and astrometric precision available would prevent us from reaching useful constraints on the orbits.

\begin{figure}
\epsscale{1}
\plotone{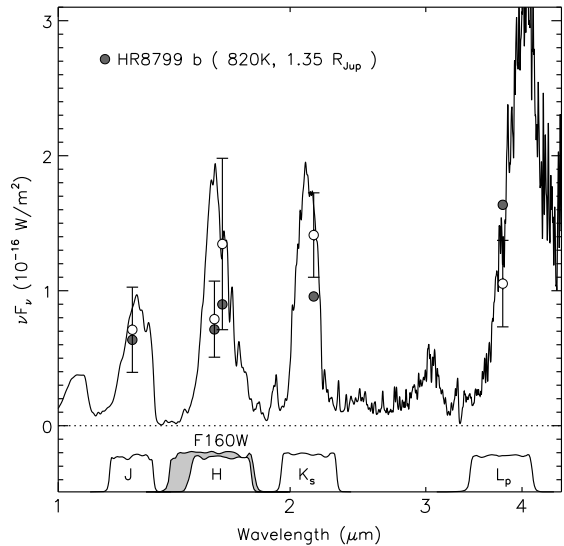}
\caption{\label{fig:sed} Synthetic spectrum from a 820~K, $\log{g} = 4.0$ model atmosphere with vertically stratified clouds. Filled symbols are the band-integrated fluxes and observations are shown as open symbols with 3$\sigma$ error bars.  Filter response curves are plotted below the spectrum.}
\end{figure}

The low luminosity and red near-infrared color of HR~8799~b ($J-K_{\rm s} = 2.25$) are indicative of atmospheric dust, but constraining the dust properties (composition and distribution) requires a detailed comparison of its SED with various model atmosphere predictions. Figure~\ref{fig:sed} compares the new F160W photometric point, along with earlier $J$, $H$, $K_{\rm s}$ and $L^\prime$-band photometry, to an intermediate (vertically stratified) cloud model (Barman et al., in prep.) with parameters selected by comparing the planet's age and luminosity to substellar evolution tracks \citep{marois08}. The observed and model photometry are listed in Table~\ref{tbl:phot}. The F160W filter is wide enough to incorporate a substantial fraction of the water band between the $J$ and $H$ bands. Since the depths of the water bands are greatly reduced as dust content increases, this filter, and other NICMOS band-passes, have greater potential for constraining the atmospheric dust content than the standard ground-based near-IR windows.   As can be seen in Fig.~\ref{fig:sed} (and Table~\ref{tbl:phot}), the new F160W photometric point is in excellent agreement with this intermediate cloud model, and about 2$\sigma$ fainter than an extreme dusty atmosphere having clouds that blanket most of the atmosphere. While the overall agreement with the extreme cloud atmosphere model appears to be relatively good, based on Table~\ref{tbl:phot}, this model requires $T_{\rm eff} = 1600$~K (cf $\sim$800~K for the intermediate cloud model) and a very small radius ($\sim$4 Earth radii) to match the observed luminosity.  Both of which are inconsistent with formation and evolution models.

\section{Concluding remarks}

The work presented here demonstrates that the LOCI algorithm can take HST data a step further, extending the sensitivity to planets beyond what has been achieved before. The good performance of the LOCI algorithm with HST data is due in large part to the high stability of the optical aberrations of HST and the consistent, accurate pointing between the different visits. Over the years, a large archive of HST observations aimed at detecting exoplanets has been assembled. Based on the present results, it would be interesting to look again at this set of data, using the LOCI algorithm, to see if any other sources have been missed by previous searches. On a related note, it could be interesting to attempt a re-analysis of archival ground-based AO observations using an approach similar to that used here. However, this could turn out be much less efficient than for HST given the larger evolution of the PSF structure for ground-based telescopes.

Looking into the future, the approach presented here should definitely be an integral part of the strategy for the search and characterization of planets with future space telescopes, such as the James Webb Space Telescope (JWST) or a Terrestrial Planet Finder. JWST, for instance, is expected to be relatively stable in temperature and should not suffer from the breathing problem experienced by HST. Its PSF should therefore be more stable than that of HST and the LOCI algorithm will likely perform extremely well, especially for observations obtained within a given wavefront adjustment campaign, typically every 14 days \citep{gardner06}. A preliminary analysis of the JWST PSF temporal evolution done by \citet{makidon08} indeed suggests that PSF subtraction should perform well. For this approach to work, enough attention should be paid to the accurate positioning of the different targets, in particular for observations made with the coronagraphic occulting masks. Having an efficient PSF subtraction strategy will be absolutely necessary for JWST given its complicated, speckled PSF structure arising from its segmented pupil, in addition to the speckles arising from the unavoidable optical aberrations. While roll-deconvolution could perform well for JWST in the absence of telescope breathing, the LOCI approach remains very interesting given the observatory's maximum roll angle of only $\pm5\degr$, which will severely limit the use of roll-deconvolution at small separations.

\acknowledgments

The authors would like to thank Yanqin Wu and Andrew Shannon for discussions of the possible orbits of this system, Bruce Macintosh for discussions relating to the new NICMOS PSF subtraction approach used here, and Markus Janson for discussions about various aspects of the NICMOS data. D.L. and C.M. are supported in part through postdoctoral fellowships from the Fonds Qu\'eb\'ecois de la Recherche sur la Nature et les Technologies. R.D. is supported through a grant from the Natural Sciences and Engineering Research Council of Canada.


\begin{thebibliography}{15}
\expandafter\ifx\csname natexlab\endcsname\relax\def\natexlab#1{#1}\fi

\bibitem[{{Fabrycky} \& {Murray-Clay}(2008)}]{fabrycky08}
{Fabrycky}, D.~C., \& {Murray-Clay}, R.~A. 2008, ArXiv e-prints, astro-ph/0812.0011

\bibitem[{{Gardner} {et~al.}(2006){Gardner}, {Mather}, {Clampin}, {Doyon},
  {Greenhouse}, {Hammel}, {Hutchings}, {Jakobsen}, {Lilly}, {Long}, {Lunine},
  {McCaughrean}, {Mountain}, {Nella}, {Rieke}, {Rieke}, {Rix}, {Smith},
  {Sonneborn}, {Stiavelli}, {Stockman}, {Windhorst}, \& {Wright}}]{gardner06}
{Gardner}, J.~P. et al. 2006, Space Science Reviews, 123, 485

\bibitem[{{Gray} \& {Kaye}(1999)}]{gray99}
{Gray}, R.~O., \& {Kaye}, A.~B. 1999, \aj, 118, 2993

\bibitem[{{Kalas} {et~al.}(2008){Kalas}, {Graham}, {Chiang}, {Fitzgerald},
  {Clampin}, {Kite}, {Stapelfeldt}, {Marois}, \& {Krist}}]{kalas08}
{Kalas}, P. et al. 2008, Science, 322, 1345

\bibitem[{{Krist}(1993)}]{krist93}
{Krist}, J. 1993, in ASP Conf. Series,
  ed. R.~J. {Hanisch}, R.~J.~V. {Brissenden}, \& J.~{Barnes}, Vol.~52, 536

\bibitem[{{Lafreni{\`e}re} {et~al.}(2007{\natexlab{a}}){Lafreni{\`e}re},
  {Doyon}, {Marois}, {Nadeau}, {Oppenheimer}, {Roche}, {Rigaut}, {Graham},
  {Jayawardhana}, {Johnstone}, {Kalas}, {Macintosh}, \&
  {Racine}}]{lafreniere07b}
{Lafreni{\`e}re}, D. et al. 2007{\natexlab{a}}, \apj, 670, 1367

\bibitem[{{Lafreni{\`e}re} {et~al.}(2007{\natexlab{b}}){Lafreni{\`e}re},
  {Marois}, {Doyon}, {Nadeau}, \& {Artigau}}]{lafreniere07a}
{Lafreni{\`e}re}, D. et al. 2007{\natexlab{b}}, \apj, 660, 770

\bibitem[{{Lafreni{\`e}re} {et~al.}(2008){Lafreni{\`e}re}, {Jayawardhana}, \&
  {van Kerkwijk}}]{lafreniere08}
{Lafreni{\`e}re}, D. et al. 2008, \apjl, 689, L153

\bibitem[{{Lagrange} {et~al.}(2008){Lagrange}, {Gratadour}, {Chauvin}, {Fusco},
  {Ehrenreich}, {Mouillet}, {Rousset}, {Rouan}, {Allard}, {Gendron}, {Charton},
  {Mugnier}, {Rabou}, {Montri}, \& {Lacombe}}]{lagrange08}
{Lagrange}, A.~. et al. 2008, ArXiv e-prints, astro-ph/0705.4290

\bibitem[{{Lowrance} {et~al.}(2005){Lowrance}, {Becklin}, {Schneider},
  {Kirkpatrick}, {Weinberger}, {Zuckerman}, {Dumas}, {Beuzit}, {Plait},
  {Malumuth}, {Heap}, {Terrile}, \& {Hines}}]{lowrance05}
{Lowrance}, P.~J. et al. 2005, \aj, 130, 1845

\bibitem[{Makidon {et~al.}(2008)Makidon, Sivaramakrishnan, Soummer, Anderson,
  \& van~der Marel}]{makidon08}
Makidon, R.~B. et al. 2008, in Proc. SPIE, 7010, 70100O

\bibitem[{{Marois} {et~al.}(2006){Marois}, {Lafreni{\`e}re}, {Doyon},
  {Macintosh}, \& {Nadeau}}]{marois06}
{Marois}, C. et al. 2006, \apj, 641, 556

\bibitem[{{Marois} {et~al.}(2008){Marois}, {Macintosh}, {Barman}, {Zuckerman},
  {Song}, {Patience}, {Lafreniere}, \& {Doyon}}]{marois08}
{Marois}, C. et al. 2008, Science, 322, 1348

\bibitem[{{Royer} {et~al.}(2007){Royer}, {Zorec}, \& {G{\'o}mez}}]{royer07}
{Royer}, F., {Zorec}, J., \& {G{\'o}mez}, A.~E. 2007, \aap, 463, 671

\bibitem[{{Schneider} \& {Silverstone}(2003)}]{schneider03}
{Schneider}, G., \& {Silverstone}, M.~D. 2003, in Proc. SPIE, 4860, 1

\end{thebibliography}
\end{document}